\documentclass[fleqn,10pt]{wlscirep}
\usepackage[utf8]{inputenc}
\usepackage[T1]{fontenc}
\usepackage{tabularray}
\usepackage{rotating}
\usepackage{multirow}
\usepackage{hyperref}
\usepackage{physics}

\providecommand{\keywords}[1]{\textbf{Mots-clés :} #1}

\UseTblrLibrary{diagbox}
\title{Supervised contrastive learning for cell stage classification of animal embryos}

\author[1,*]{Yasmine Hachani}
\author[1]{Patrick Bouthemy}
\author[1,2]{Elisa Fromont}
\author[3]{Sylvie Ruffini}
\author[3]{Ludivine Laffont}
\author[3,4]{Alline de Paula Reis}
\affil[1]{Inria center at Rennes University, France}
\affil[2]{University of Rennes, IRISA, France}
\affil[3]{Paris-Saclay University, UVSQ, INRAE, BREED, France}
\affil[4]{The National Veterinary School of Alfort (EnvA), France}

\affil[*]{yasmine.hachani@inria.fr}

\begin{abstract}
Video microscopy, when combined with machine learning, offers a promising approach for studying the early development of \textit{in vitro} produced (IVP) embryos. However, manually annotating developmental events, and more specifically cell divisions, is time-consuming for a biologist and cannot scale up for practical applications. We aim to automatically classify the cell stages of embryos from 2D time-lapse microscopy videos with a deep learning approach. We focus on the analysis of bovine embryonic development using video microscopy, as we are primarily interested
in the application of cattle breeding, and we have created a Bovine Embryos Cell Stages (ECS) dataset. The challenges are three-fold: (1) low-quality images and bovine dark cells that make the identification of cell stages difficult, (2) class ambiguity at the boundaries of developmental stages, and (3) imbalanced data distribution. To address these challenges, we introduce CLEmbryo, a novel method that leverages supervised contrastive learning combined with focal loss for training, and the lightweight 3D neural network CSN-50 as an encoder. We also show that our method generalizes well. CLEmbryo outperforms state-of-the-art methods on both our Bovine ECS dataset and the publicly available NYU Mouse Embryos dataset.
\end{abstract}

\begin{document}
\flushbottom
\maketitle
%
%
\thispagestyle{empty}

\section*{Introduction}

The analysis of embryonic development is of key interest in better understanding biological mechanisms and addressing a range of human and animal concerns.
When studying \textit{in vitro} produced (IVP) embryos, the analysis can benefit from image sequences obtained from a video microscopy setup. In this paper, we will study the embryonic developmental stages of IVP embryos from time-lapse video sequences. The normal development of a mammalian embryo is characterized by a series of mitoses in which each cell divides into two daughter cells (also known as cleavage). Successive developmental stages, or cell stages, are characterized by a given number of cells, starting from one cell. We formulate this analysis of embryonic development as a supervised classification problem. Each class corresponds to a cell stage and is related to a number of cells. This problem is unbalanced, and the vast majority of images show a number of cells of a power of two. In the sequel, we will call intermediate stages the other cell stages (3-cell, 5-cell, 6-cell, and 7-cell stages). We will not proceed with an explicit numbering of cells based on cell segmentation as done in \cite{stage-detection-yolo}, but rather to a direct classification of images based on a deep learning approach.

Observation of embryos by conventional microscopy requires them to be removed from the incubator, leading to disturbances in the temperature and pH of the culture medium that can be detrimental to embryo development\cite{NGUYEN201864}.
Video microscopy performed directly in the incubator provides a reliable way of capturing images of embryos at regular and short intervals. It enables embryologists to determine cell numbers almost in real time, without interfering with the culture conditions required for normal embryo development. It also has the advantage of being compatible with
further transfer to establish pregnancy\cite{FumieMAGATA20232022-131}. 
However, manually determining the number of cells is a time-intensive task and limits the extent to which laboratories equipped with a video microscope can make more extensive use of this information. This argues for automation.

\begin{figure*}
\centering
\resizebox{\columnwidth}{!}{
    \includegraphics{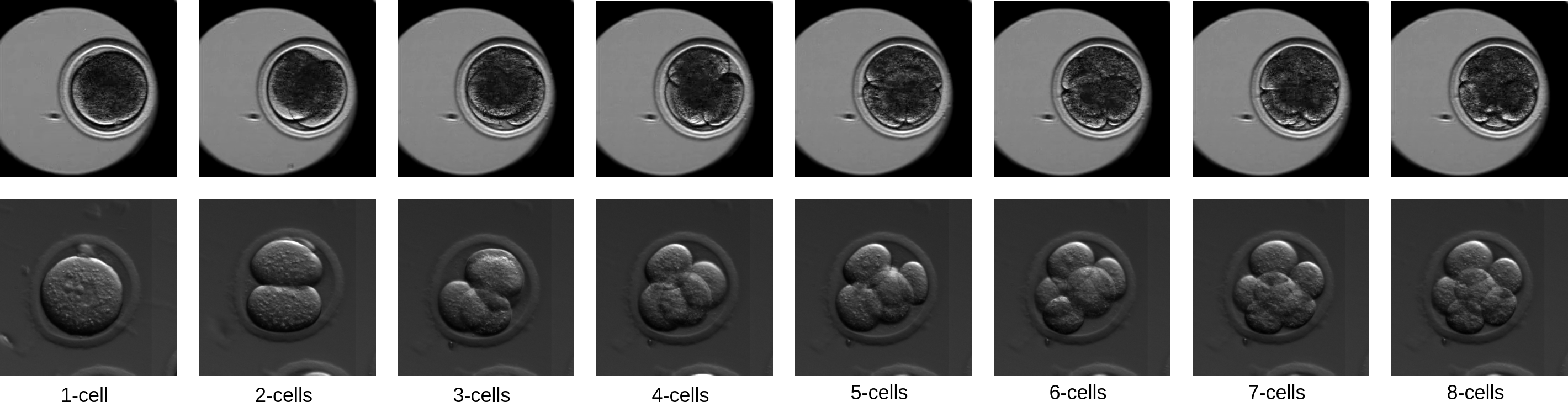}}
\caption{Sample images of the different cell stages of bovine (top row) and mouse (bottom row) embryo development. The bovine embryo only occupies a small part of the image, it is located in a micro-well (light grey) within the Petri dish (black). The bovine embryo is darker. The mouse embryo is more transparent. Both embryos are surrounded by the zona pellucida. Images are taken from the Bovine ECS dataset and from the NYU Mouse Embryo dataset respectively.}
\label{fig:embryo-dev}
\end{figure*}

We will focus on the analysis of bovine embryonic development using video microscopy, as we are primarily interested in the application for cattle breeding. Its use in the bovine domain is recent but promising. More specifically, it allows one to assess whether embryos can be transferred to a cow uterus\cite{sfr}. From an image processing perspective, images of bovine embryos are especially difficult to process due to their dark appearance, as shown in Figure \ref{fig:embryo-dev}. To demonstrate the generality of our method, we will also apply it to the analysis of mouse embryonic development.





Existing work dealing with cell stage classification and making use of deep learning techniques can be divided into two categories. The first category of work considers every frame of a time-lapse video independently and employs 2D convolutional neural networks (CNNs) to achieve the classification\cite{emca}. This is usually followed by a post-processing step that enforces a monotonic progression of the number of cells as done in \cite{stage-detection-fusion}, \cite{stage-detection-multi-task}, \cite{celldivision} and \cite{stage-detection-resnext}, all using dynamic programming to apply that constraint. In \cite{stage-detection-resnext}, the development stages are identified through several sub-steps that also take into account the rate of embryo fragmentation. In \cite{stage-detection-yolo}, the authors adopt an approach based on object detection using the YOLO-v5 network \cite{yolov5} and perform cell counting. Very recently, the DLT-Embryo method was proposed\cite{dlt-embryo}. It combines dual-branch local feature fusion (DLF) modules and transformer encoder modules to extract both local and global features from each frame. These methods make limited use of temporal information.

Embryologists take into account dynamic information to decide the completion of the cell cleavage event. In fact, the temporal context helps maintain consistency between time-lapse images.
The second category of methods integrates temporal information in the architecture, and then exploits it as soon as the training stage.

The ESOD method \cite{stage-detection-synergic} employs a 2D-CNN followed by a Long Short-Term Memory (LSTM) network classifier, and involves a synergic loss to learn embryo-independent features. The CNNs-CRF method proposed in \cite{cnn-crf} involves a conditional random field (CRF) to favor the monotonic progression of the number of cells. In addition, the neural network leverages temporal information through two streams with different input, one single frame for the first one and two concatenated consecutive frames for the second one. The R2D1 method \cite{stage-detection-r2d1-vit} considers the spatio-temporal video as a volume, uses 3D-CNN (i.e, 2D+t) to work directly on video subsequences, and applies the Viterbi algorithm to enforce a monotonic order constraint. The EmbryosFormer framework \cite{embryosformer} is the first to involve transformer architecture, building a three-headed encoder-decoder transformer inspired by the Deformable DETR model \cite{def_detr}. All of these methods were evaluated on datasets of human and mouse embryos, except R2D1, which was evaluated only on humans.


As stated before, our objective is to classify the cell stages of an animal embryo from 2D time-lapse videos acquired by video-microscopy. 
This prediction will be based on a 3D-CNN architecture and will leverage supervised contrastive learning (SCL)\cite{scl}. 
As done by the other authors, we assume that the $n$-cell stage lasts until the $n+1$-cell stage appears.
We formulate this problem as a multi-class supervised classification. Each frame of an input time-lapse video will be assigned a class from the one-cell stage to the nine-and-plus-cell stage. 

The classification problem is challenging, even if it is supervised.
The cell classes are heavily imbalanced. The intermediate cell stages are largely underrepresented. Frames located just around the end of a cell cleavage, consequently, around a stage change, are particularly hard to classify. In fact, immediately before the end of the daughter-cell separation, 
two daughter 
cells can be seen despite the incomplete membrane separation. In addition, immediately after complete separation, cell movements due to rearrangement of newly created cells in the embryo can be confused with the movements required for cell separation.
Embryo images usually exhibit low contrast, noise, and intricate motion.
Bovine embryos are harder to study using video microscopy than their human or mouse counterparts due to the darkness of their cells. Cell cleavage may last longer than for mouse embryos.

Work that combines video microscopy and machine learning for bovine embryos is rare. In\cite{bovine_embryo0}, a method was developed for the prediction of embryo transferability based on random forests. It requires very detailed manual annotations of each video. In \cite{sfr}, we have designed a 3D-CNN called SFR that includes three paths and the focal loss\cite{focal}, to achieve transferability prediction directly from videos. To our knowledge, we are the first to propose a deep learning-based model to classify the stages of early development of bovine embryos and to leverage SCL for this kind of task.

SCL \cite{scl} is a contrastive learning technique that uses labeled data to learn more robust and semantically meaningful representations. It extends the principle of contrastive learning, which aims to pull similar data points closer in the representation space and push dissimilar ones further apart. SCL incorporates label information to define similarity and dissimilarity between samples. SCL has been shown to be a promising approach in the field of medical image analysis. 
It has been applied to a variety of tasks, including medical image classification \cite{scl-histo-classif}
and segmentation \cite{medical-image-scl-seg}.


\section*{Materials and methods}
\hypertarget{sec:method}{}
This work had several objectives: 1) to propose a new method for classifying bovine embryo cell stages, 2) to verify that our model generalizes well to other mammalian species once retrained, 3) to compare its performance with four other existing methods (ESOD, CNNs-CRF, EmbryosFormer, R2D1) commented on in the Introduction section. The method was tested on our Bovine Embryos Cell Stage dataset (or Bovine ECS dataset for short) and the NYU mouse embryos dataset described below.
\subsection*{Datasets}
\subsubsection*{Bovine Embryos Cell Stage dataset}
We have acquired, as described by \cite{bovine_embryo0}, a video dataset of IVP bovine embryos that we call the Bovine ECS dataset. The embryo production process, the acquisition of embryo videos and the main features of our dataset are described in this section.
The bovine embryos were obtained from oocytes recovered \textit{post mortem} on slaughterhouse ovaries, fertilized \textit{in vitro} and individually placed in the incubator in 16-microwell Primovision dishes filled with SOF medium (Synthetic Oviduct Fluid) covered with oil and cultured for seven or eight days at 38.5°C in a 5\%CO\textsubscript{2}, 5\%O\textsubscript{2} and 90\%N\textsubscript{2} humidified atmosphere. Images of the Petri dishes were automatically taken every fifteen minutes throughout the embryo culture (from 22 to 168 or 192 hours of development, depending on the number of days of cultivation) by a PrimoVision system equipped with a transmission light microscope. The resulting 2D time-lapse videos were subsequently divided into sixteen individual time-lapse videos. 
The different cell stages of each embryo were then annotated by an embryologist. The first image presenting two daughter cells and their entire membranes was considered as a stage change. These annotations were used for training and evaluation.
The videos used for this study comprise 300 frames of 256 x 256 pixels showing one unique embryo. Knowing that the first frame was taken at the beginning of the embryo culture (at $t_0=22h$), and that the interval between two successive images was 15 minutes, the end of the 300-frame video corresponds to the fourth day of embryo development and between 8 and 16 cells for normal development in bovine species.
Our final data set consists of 1221 videos, of which 854 were taken for training our models, 122 were for parameter validation, and 245 videos for evaluating the performance (test). Details on the data distribution of each cell stage in the dataset are given in Table \ref{tab:data-distrib}.

\subsubsection*{Mouse Embryos dataset}

We also consider mouse embryos to evaluate the generality of our proposed method. The NYU Mouse Embryo dataset\cite{mouse-dataset} contains 100 videos of developing mouse embryos, originally created for the task of cell tracking. They were acquired with a Nikon Eclipse Ti inverted microscope and a heated stage-top Tokai incubator. Each time-lapse video, which shows one unique embryo, consists of 480 × 480 pixels images captured every 7 minutes, resulting in 314 images per sequence on average. The image capture frequency is higher for mouse embryos than for bovine ones, as mouse embryos develop faster. Each video ends in the 8-cell stage. After downloading the dataset, we had access to 99 videos, divided into 72 sequences for training, 8 for validation, and 19 for testing. We used the annotations provided by the authors of the CNNs-CRF\cite{cnn-crf} method.
Details on the data distribution of each cell stage in the datasets are also given in Table \ref{tab:data-distrib}.

\begin{table}
\centering
\caption{Distribution of cell stages computed over all videos (\textit{i.e}, all embryos) of the Bovine ECS and Mouse Embryos datasets, used for training, validation and test. We discarded the 8-cell stage frames of the Mouse Embryos dataset as they appear only in the last two frames of each video, making it extremely underrepresented and with a high risk to bias the model training phases.}
\begin{tabular}{c|ccc|c||c|ccc|c} 
\hline
\multicolumn{5}{c||}{Bovine}                                                          & \multicolumn{5}{c}{Mouse}                            \\ 
\hline
Cell stage         & Train           & Val            & Test            & Total            & Cell stage    & Train & Val & Test          & Total       \\ 
\hline
1             & 29279           & 4332           & 8393            & 42004            & 1         & 3496 & 485 & 998           & 4979        \\
2             & 22046           & 3467           & 6168            & 31681            & 2         & 11971 & 1311 & 3090          & 16372       \\
3             & 4855            & 557            & 1579            & 6991             & 3         & 567  & 49  & 138           & 754        \\
4             & 18120           & 2537           & 4912            & 25569            & 4         & 6353 & 715 & 1689          & 8757        \\
5             & 1560            & 222            & 556            & 2338             & 5         & 354  & 31  & 76           & 461        \\
6             & 2549            & 155            & 612            & 3316             & 6         & 472  & 54  & 155           & 681        \\
7             & 1531            & 315            & 391            & 2237             & 7         & 620  & 39  & 96           & 755        \\
8             & 34647           & 4632           & 8353            & 47632            & \multirow{2}{*}{} & \multicolumn{3}{c|}{\multirow{2}{*}{}} & \multirow{2}{*}{} \\
9+             & 7981            & 1603           & 2386            & 11970            &          & \multicolumn{3}{c|}{}         &          \\ 
\hline
\multicolumn{1}{l|}{Total} & \multicolumn{1}{l}{122568} & \multicolumn{1}{l}{17820} & \multicolumn{1}{l|}{33350} & \multicolumn{1}{l||}{173738} & Total       & 23833 & 2684 & 642           & 32759       
\end{tabular}
\label{tab:data-distrib}
\end{table}

\subsection*{Description of our method (CLEmbryo)}
We propose an original method, called CLEmbryo, to classify the cell stages that appear during embryonic development. It integrates three key components: (1) a modified version of the SCL framework, (2) a loss function that combines focal loss (FL) and supervised contrastive loss (called SupCon\cite{scl}), and (3) the Channel-Separated Convolutional Network (CSN)\cite{CSN} 3D-CNN architecture as an encoder.

\subsubsection*{SCL framework}
\begin{figure}[t!]
\centering
\resizebox{\columnwidth}{!}{
    \includegraphics{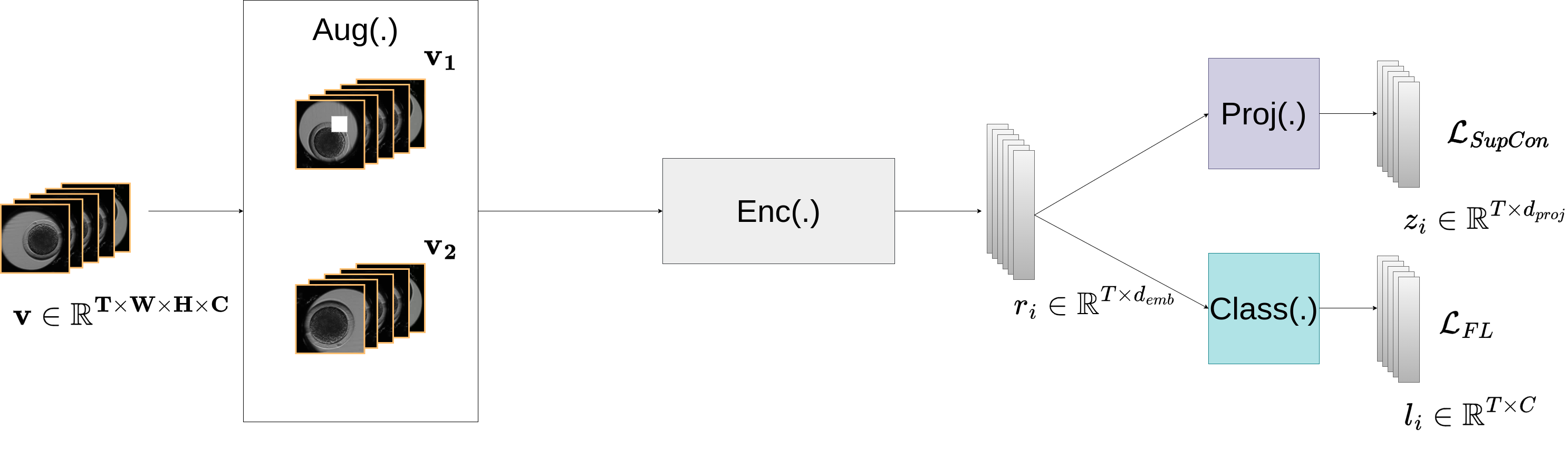}}
  \caption{The SCL framework of our CLEmbryo method with the data augmentation module $Aug(.)$, the encoder network $Enc(.)$, the projection network $Proj(.)$ and the classification network $Class(.)$. The SupCon loss $\mathcal{L}_{SupCon}$ is computed using the output of $Proj(.)$, while the focal loss $\mathcal{L}_{FL}$ is computed with the output of $Class(.)$. The data augmentation module generates two augmented version of the original sequence, and each version is processed independently. We set $d_{proj}=128$ in all the experiments. The value of $d_{emb}$ depends on the encoder CNN and is set to 512 when CSN-50 is involved.}
  \label{fig:clembryo}
\end{figure}

Inspired by the original paper on supervised contrastive learning\cite{scl}, our framework consists of four components: the data augmentation module $Aug(.)$, the encoder network $Enc(.)$, the projection network $Proj(.)$ and the classification network $Class(.)$.
The overall framework is presented in Figure \ref{fig:clembryo}. In contrast to recent implementations of the SCL framework with two losses and a single training stage, we also carry out a single training stage with two losses, but corresponding to two heads, classification head, and projection head.

At training time, each video subsequence $\textbf{v}\in\mathbb{R}^{T\times H\times W\times C}$, where $H$ and $W$ respectively denote the height and width of every image, $T$ the temporal length of the subsequence, and $C$ the number of image channels ($C=1$), passes through the data augmentation module that generates two random augmentations $\textbf{v}_i\in\mathbb{R}^{T\times H\times W\times C}$ ($i=1,2$) representing different views with distinct information derived from the original input. The list of the augmentations taken into account is provided in the dedicated section below. Then, each view $\textbf{v}_i$ is processed independently.

As illustrated in Fig.\ref{fig:clembryo}, the encoder embeds each frame $v_{i,t}$ of each subsequence $\textbf{v}_{i}$ in a vector, $Enc(v_{i,t}) = \textbf{r}_{i,t} \in \mathbb{R}^{d_{emb}}$, $t\in \mathopen{[}1, T\mathclose{]}$. Every representation vector is normalized to the unit hypersphere in $\mathbb{R}^{d_{emb}}$. The projection head and the classification head process the representation vectors simultaneously. The projection head maps $\textbf{r}_{i,t}$ to a vector $\textbf{z}_{i,t} = Proj(\textbf{r}_{i,t})\in\mathbb{R}^{d_{proj}}$ to reduce the number of operations required to compute the SupCon loss. $Proj(.)$ is a single linear layer and all projection vectors are normalized to the unit hypersphere in $\mathbb{R}^{d_{proj}}$. Likewise, the classification head is a single linear layer that maps $\textbf{r}_{i,t}$ to a classification vector $\textbf{l}_{i,t} = Class(\textbf{r}_{i,t})\in\mathbb{R}^{K}$, with $K$ denoting the number of classes from which the focal loss is computed. 
Once training is achieved, both the data augmentation module and the projection head are removed, and inference is performed on the resulting architecture.

\subsubsection*{Loss function}
The training loss involves two loss terms as illustrated in Figure \ref{fig:clembryo}, the SupCon loss and the classification loss.
The SupCon loss is built on the concept of contrastive loss extended to supervised scenarios by the authors of\cite{scl}. This loss adds information on whether two elements belong to the same class, thus helping to fix the limitations of self-supervised contrastive learning, which might push apart the representations of same-class samples. By enforcing clustering within classes, the SupCon loss ensures that the encoder produces representations that are more robust and discriminative, ultimately to the benefit of downstream classification tasks. Let us consider N video subsequences. The data augmentation module generates two augmented versions of each subsequence. We note $\mathcal{J} = \{1,...,N\}$ the set of indexes $j$ of all subsequences $\textbf{v}^{j}_{i}$ and $\textbf{y}^{j}_{i}\in[1,K]^T$ their associated vector of class labels. For a given frame $v^{j}_{i, t}$, let us also consider $A(j,i,t) = \mathcal{J}\times \mathopen{[}1, 2\mathclose{]} \times \mathopen{[}1, T\mathclose{]} \setminus \{(j,i,t)\}$, the set of indexes of all the augmented frames except $v^{j}_{i,t}$ and $P(j, i, t) = \{(q, p, s) \in A(j, i, t), y^{q}_{p, s}=y^{j}_{i,t}\}$, the positive set of $v^{j}_{i, t}$. $P(j, i, t)$ contains the other augmented version of the original frame $v^{j}_{t}$, like purely self-supervised approaches, but also all augmented frames with the same label as $v^{j}_{i, t}$.
The SupCon loss writes:
\begin{equation}
\mathcal{L}_{SupCon} = \sum_{j\in \mathcal{J}}\sum_{i=1}^{2}\sum_{t=1}^{T}\mathcal{L}_{SupCon}(v^{j}_{i, t},y^{j}_{i,t}) = \sum_{j\in \mathcal{J}}\sum_{i=1}^{2}\sum_{t=1}^{T} -\frac{1}{|P(j,i,t)|}\sum_{(q, p, s)\in P(j,i,t)}\log(\frac{\exp(\frac{\textbf{z}^{j}_{i, t} \cdot \textbf{z}^{q}_{p, s}}{\tau})}{\sum_{(\beta, \delta, \nu)\in A(j,i,t)}\exp(\frac{\textbf{z}^{j}_{i, t} \cdot \textbf{z}^{\beta}_{\delta, \nu}}{\tau})}),
\end{equation}
with $\textbf{z}^{j}_{i,t} = Proj(Enc(v^{j}_{i,t}))$, and $\tau\in\mathbb{R}_{+}^{\ast}$ a scalar temperature parameter. 
As in \cite{multimodal-scl}, this loss can be used as an auxiliary objective during the training process.

We could adopt different loss functions for the classification objective. Since our data are unbalanced with respect to cell stages, we have adopted the focal loss\cite{focal}, initially introduced for the object detection task. The focal loss helps to mitigate this imbalance, while focusing on the most difficult examples. The focal loss writes:
\begin{equation}
\mathcal{L}_{FL} = \sum_{j \in \mathcal{J}}\sum_{i=1}^{2}\sum_{t=1}^{T}\mathcal{L}_{FL}(v_{i,t}^j,y_{i,t}^j) = \sum_{j \in \mathcal{J}}\sum_{i=1}^{2}\sum_{t=1}^{T}-\mathopen{[}\sum_{k=1}^{K}\alpha_k(1-\hat{p}(k|v_{i,t}^j))^\gamma p(k|v_{i,t}^j)\log\hat{p}(k|v_{i,t}^j) \mathopen{]},
\end{equation}

where $k$ denotes one of the K classes, $\hat{p}(k|v^{j}_{i,t})$ the predicted probability of having class $k$ given an augmented frame $v^{j}_{i,t}$, and $p(k|v^{j}_{i,t})$ the true one. The latter is equal to $1$ if the right class $k$ is assigned to $v^{j}_{i,t}$, since we are dealing with supervised classification. In addition, $\alpha_k$ is the weight for class $k$, $\gamma$ the focusing parameter. The larger $\gamma$, the less importance is given to well-classified samples.

Our final objective function combines the proposed SupCon loss for representation learning and the focal loss for classification as follows:
\begin{equation}
\mathcal{L}_{train} = w\mathcal{L}_{SupCon} + (1-w)\mathcal{L}_{FL}.
\end{equation}
\noindent In all experiments, we set $w=0.5$, $\tau=0.5$ and $\gamma=2$, this value for $\gamma$ being recommended in \cite{berntsen2022} and \cite{sfr}.

\subsubsection*{Encoder neural network}
We believe that a 3D-CNN network is more adapted to properly capture the spatio-temporal features characterizing embryonic development than a recurrent neural network as previously demonstrated \cite{sfr}, \cite{stage-detection-r2d1-vit}. 
By 3D, we mean two spatial dimensions and the temporal one (2D+t). The 3D-CNN selected for the encoder must be accurate and lightweight as 3D convolutions are computationally costly. For this reason, we chose the CSN \cite{CSN}, which effectively meets these requirements. The CSN architecture leverages group convolution to separate channel interactions from spatio-temporal interactions. This significantly reduces computational cost while preserving sufficient channel interactions to maintain high accuracy. We use CSN-50 with 50 layers and 13.6M parameters. For comparison, 3D-ResNet-50 approximately comprises 45M parameters. We have adapted the CSN architecture to avoid any reduction in time resolution. To ensure that a prediction is made for each frame in the sequence, we set the pooling and stride to one in the temporal dimension.

\subsubsection*{Data augmentation}
\label{subsubsec:data-aug}
Data augmentation should introduce sufficient diversity in the versions of the original input without altering the biological meaning of the videos.
Each augmentation is selected at random and is applied to
all images of a given sequence. The set of augmentations includes random change in brightness and contrast, horizontal and vertical flipping, random rotation from 0 to 30 degrees, random translation, random cropping, and cutout\cite{cutout} that occludes a random square region of the input video.

\subsection*{Implementation details}

\subsubsection*{Implementation of CLEmbryo}
CLEmbryo was trained using the AdamW optimizer \cite{adamw}, with a learning rate of $5\times10^{-4}$ and the other parameters kept at their default values. We applied a learning rate reduction by a factor of 0.1 in the plateau scheduler. During training,
a subsequence of consecutive frames is randomly selected from each video at each epoch.
We prefer to use as input a subsequence rather than a single frame, as local temporal information is beneficial to classify the cell stage.
As for the size of the subsequence, the larger the time window, the more temporal context is provided, but this leads to higher computational costs. We set the size at ten, this choice being motivated in the ablation study. We trained the models using mini-batches of 32 samples in total, meaning that within the SCL framework, 16 different subsequences are randomly selected. We applied the stochastic weight averaging technique (SWA)\cite{swa}, which improves the generalization of our models by averaging the network weights obtained at several well-chosen epochs. We used early stopping to end training, when the loss computed on the validation set increases ten epochs in a row. Then, we selected the model at the epoch with the lowest loss in the validation set.

During inference, we apply temporal averaging implemented with an overlapping rolling window technique on ten consecutive frames and a step of four. This means that for frames not located at the beginning or the end of the video sequence, their classification vector is computed as the mean of three classification vectors issued from three different windows.


\subsubsection*{Implementation of the other methods}

We used the code available online and the parameters provided by the respective authors for the other methods. We only changed the optimizer to AdamW, which is known to handle L2 regularization better than Adam. The hyperparameters are set to the values reported as optimal by the authors. 
During training, R2D1 uses a batch size of 32 distinct subsequences of ten consecutive frames each. For EmbryosFormer, a 2D-ResNet-50 is pre-trained using the SimCLR framework \cite{simCLR}. SimCLR has proven to be a good pre-training method to learn efficient representations, including for biomedical images\cite{simCLR-pretraining}. The transformer model is then trained using the extracted feature maps as input, with a learning rate of 1e-4, a batch size of 32 sequences and a warm-up phase followed by a cosine decay scheduler. All feature maps from the 2D-CNN of all frames in a video are used as input. In CNNs-CFR, a warm-up phase is followed by a learning rate reducer scheduling. The learning rate is set to 1e-4 and the batch size to 4 sequences, each consisting of 50 randomly selected consecutive frames. The model has two inputs: a single frame for the main path and two consecutive frames for the temporal path. ESOD uses a learning rate of 1e-4, a learning rate reduction in the plateau scheduler and MixUp\cite{mixup2018}. It processes two sequences of 32 randomly selected images and compares these sequences using a synergic loss model with dual input.

In contrast to our approach, the four methods with which we compared our method, use a post-processing step or a mechanism to enforce a monotonic constraint on the predicted labels at test time. R2D1 uses the Viterbi algorithm to ensure temporal consistency. Embryosformer integrates this constraint via its dedicated segmentation head. CNNS-CRF uses its CRF to refine the predictions accordingly. ESOD directly exploits the output of its LSTM layers.

\subsection*{Evaluation metrics}
To evaluate the performance of all methods, we consider three metrics: global accuracy, F1-score per class, and temporal accuracy. 
Global accuracy $\vartheta$ is the ratio of images correctly classified by the model, and is given by $\text{N}_{corr}/\text{N}_{total}$, where $\text{N}_{corr}$ is the number of correctly classified images and $\text{N}_{total}$ the total number of images. The F1-score per class is defined as the harmonic mean of precision (P) and recall (R) per class: $\text{F1} = 2 \text{P} \text{R} / (\text{P} + \text{R})$.
Precision (P) and recall (R) are respectively defined by $\text{P} = \frac{\text{TP}}{\text{TP} + \text{FP}}$, and $\text{R} = \frac{\text{TP}}{\text{TP} + \text{FN}}$, where TP, TN, FP, and FN represent the number of true positives, true negatives, false positives, and false negatives, respectively.
The F1-score provides a balanced measure of the method performance.
Introduced by R2D1\cite{stage-detection-r2d1-vit}, the temporal accuracy $\zeta$ is defined as the average proportion of cell stage transitions predicted within a specified time range of the corresponding actual transitions. In this context, the time difference between the predicted and actual transitions is considered acceptable if it is less than a threshold that we set to two frames. This threshold accounts for variability in label annotations by different expert biologists.
We have: 
$\zeta = \frac{\theta-\theta_{far}}{\theta}$,
where $\theta$ is the total number of cell stage transitions and $\theta_{far}$ refers to the number of transitions predicted more than two frames away from the ground truth.

\section*{Results}
\hypertarget{sec:results}{}

The results of the ablation study performed on the Bovine ECS dataset are presented in Tables \ref{tab:ablation} and \ref{tab:seq-length}. The results of the comparison of the ESOD, CNNs-CRF, EmbryosFormer, R2D1 and our CLEmbryo with or without temporal averaging (CLEmbryo\_NTA) on the Bovine ECS and the NYU mouse datasets are presented in Tables \ref{tab:bovine-classif} and \ref{tab:mouse-classif}, respectively. The results obtained on bovine or mouse embryos with pre-training on the mouse or bovine embryos, respectively, are presented in Table \ref{tab:pretraining}.

\begin{table}[ht]
\centering
\caption{Ablation study on our Bovine ECS dataset in terms of per-class F1-score, global accuracy, and temporal accuracy. We changed one component of our model
at a time: the R(2+1)D-18 architecture instead of CSN-50, the cross-entropy loss instead of the focal loss, and our model without SCL. We carried out five evaluations each time with different training seeds, and we provide the mean and standard deviation. The best scores are highlighted in bold. All the experiments were run with a subsequence length of ten ($T=10$).}
\label{tab:ablation}
\begin{tabular}{cc|ccc|c} 
\hline
\multicolumn{2}{l|}{\diagbox{Metrics}{Ablation}} & R(2+1)D-18     & Cross-entropy     & w/o SCL   & Full CLEmbryo        \\ 
\hline
\multirow{9}{*}{F1-score per class} & 1        & 93.39 ± 0.40 & \textbf{93.72 ± 0.33} & 92.44 ± 0.47 & 93.67 ± 0.58      \\
                  & 2        & 85.50 ± 1.00 & 86.16 ± 0.93 & 84.60 ± 0.38 & \textbf{86.67 ± 0.77}      \\
                  & 3        & 24.52 ± 10.45 & 34.82 ± 6.22 & 23.29 ± 9.44 & \textbf{37.94 ± 5.23}     \\
                  & 4        & 74.82 ± 0.82 & 75.89 ± 0.77  & 74.44 ± 1.38 & \textbf{76.64 ± 1.77} \\
                  & 5        & 6.42 ± 3.94  & 14.48 ± 3.60 & 15.90 ± 6.68 & \textbf{19.76 ± 2.14}      \\
                  & 6        & 8.50 ± 9.14  & 11.55 ± 11.15 & 7.91 ± 7.06 & \textbf{15.15 ± 7.43} \\
                  & 7        & 14.26 ± 3.26 & 22.16 ± 8.91 & 24.00 ± 6.46 & \textbf{28.38 ± 6.04}      \\
                  & 8        & 85.68 ± 0.88 & \textbf{86.61 ± 3.75} & 84.52 ± 0.87 & 85.46 ± 1.43      \\
                  & 9+       & 66.09 ± 6.75 & 69.77 ± 2.15 & 68.33 ± 5.23 & \textbf{71.96 ± 2.42}      \\ 
\hline
\multicolumn{2}{c|}{Accuracy ($\vartheta$)}      & 80.52 ± 0.72  & 80.58 ± 0.51 & 79.53 ± 0.69 & \textbf{81.25 ± 0.89}      \\ 
\hline
\multicolumn{2}{c|}{Temporal accuracy ($\zeta$)}   & 53.27 ± 4.75  & 57.60 ± 4.40     & 55.61 ± 4.55 & \textbf{61.35 ± 3.43} \\ 
\hline
\multicolumn{2}{c|}{Number of parameters}       & 31.5M     & 13.6M         & 13.6M    & 13.6M         \\
\hline
\end{tabular}
\label{tab:ablation}
\end{table}

\begin{table}[th!]
\centering
\caption{Results on our Bovine ECS dataset, in terms of global accuracy and temporal accuracy, obtained by our CLEmbryo method trained with input sequence lengths of 3, 5, and 10 frames. We carried out five evaluations each time with different training seeds, and we provide the mean and standard deviation. The best scores are highlighted in bold.}
\begin{tabular}{l|lll} 
\hline
              & \multicolumn{1}{c}{3} & \multicolumn{1}{c}{5} & \multicolumn{1}{c}{10} \\ 
\hline
Accuracy ($\vartheta$)   & 75.60 ± 1.56     & 79.26 ± 1.04     & \textbf{81.25 ± 0.89}  \\
Temporal accuracy ($\zeta$) & 32.30 ± 2.41     & 47.42 ± 3.95      & \textbf{61.35 ± 3.43}  \\
\hline
\end{tabular}
\label{tab:seq-length}
\end{table}

\subsection*{Ablation study}
We carried out an ablation study on three components of our method: the architecture, the loss function and the use or not of SCL. We changed one component of our model at a time. All results are collected in Table \ref{tab:ablation}, and all models were trained and evaluated under the same conditions. First, we conducted an ablation experiment on the encoder neural network. We compared CSN-50 and R(2+1)D-18\cite{r2d1} architectures, with the latter achieving better results than a standard 3D-ResNet-18, as demonstrated in \cite{stage-detection-r2d1-vit}. All evaluation metrics, except the F1-score for the 8-cell stage, display higher values with CSN-50, justifying the choice of this CNN. In addition, CSN-50 is significantly lighter (13.6M parameters versus 31.5M). Incidentally, a transformer module is even heavier, for instance, it contains 41M parameters in the EmbryosFormer method.
The second ablation study aimed to verify whether the focal loss leads to better performance than cross-entropy. These two losses are expected to behave differently.
The former should enhance the F1-scores on intermediate classes,
improving the temporal accuracy. The latter is expected to provide better F1-scores on the most represented classes.
In our context, the behavior of the focal loss is \textit{a priori} more desirable, as we value the detection of all cell stages. 
As expected, the model trained with cross-entropy achieves better results for the predominant classes, that is, 1-cell and 8-cell classes. All other metrics, particularly temporal accuracy, are significantly improved when using the focal loss, which is why we chose this loss function.
Our last ablation experiment investigated whether the supervised contrastive learning (SCL) framework improves performance. Training CSN-50 within our supervised contrastive learning framework yields accuracy and temporal accuracy gains of 1.72 and 5.74 respectively.

Finally, we investigated the influence of the length of the input sequence on the performance of our CLEmbryo method. We trained our method with input sequence lengths of 3, 5, and 10 frames. We set the maximum sequence length to ten frames to allow a large enough batch size, which is very important for the contrastive learning-based method to converge\cite{scl}. The results are reported in Table \ref{tab:seq-length}. Global accuracy and temporal accuracy increase as the temporal window widens. This highlights the importance of providing sufficient temporal context to the contrastive learning-based method to converge.

\subsection*{Classification results}
We carried out comparative experiments on the classification of the developmental stages of bovine and mouse embryos. We performed the cell stage classification with our CLEmbryo method and four existing state-of-the-art methods, including EmbryosFormer\cite{embryosformer}, R2D1\cite{stage-detection-r2d1-vit}, CNNs-CRF\cite{cnn-crf} and ESOD\cite{stage-detection-synergic}. We were unable to compare our method with DLT-embryo as we did not have access to their code. All methods were evaluated on the same data split with five evaluations performed using the same set of training seeds. 
Our CLEmbryo method outperforms the four other methods for both datasets. It achieves higher scores for all metrics and all stages except the 8-cell stage for the Bovine ECS dataset and for all metrics and all stages for the Mouse Embryos dataset.

The results of all the compared methods obtained on the Bovine ECS dataset are presented in Table \ref{tab:bovine-classif}. This dataset is highly unbalanced, making the correct classification of the intermediate stages difficult. This explains the low scores obtained for these classes. CLEmbryo achieves better performance for all the F1-scores per class, except the 8-cell class, where R2D1 performs better. CLEmbryo reaches a global accuracy of 81.25\% with high stability (standard deviation of 0.89). The temporal accuracy of our method is also considerably better with a gain of 22.53 points compared to the second-best score obtained by R2D1. ESOD could not handle the class imbalance at all and is unable to make a right prediction of the intermediate classes. In addition, the version of our method without temporal averaging (named CLEmbryo\_NTA) still performs better than the four other methods.

From Table \ref{tab:mouse-classif}, we can observe that the Mouse Embryos dataset is easier to work with than the bovine one. Indeed, the global accuracy obtained by our CLEmbryo method is 97.73\% for the mouse 
dataset, while it was 81.25\% for the bovine dataset. 
CLEmbryo has the best results for all evaluation metrics for the mouse dataset.
We did not include the mouse 8-cell stage as
this dataset is very specific for this cell stage. The 8-class stage occurs only and systematically in the last two frames of each video. Then, this cell stage is largely under-represented and this constant configuration can bias training. 
Regarding EmbryosFormer, the results we obtained are significantly lower than those published by the authors in their paper\cite{embryosformer} (global accuracy of 98.4\%). This may be due to the way the 2D-ResNet-50 module was pre-trained, and to the choice of data split. 
In addition, the authors reported results obtained from one single experiment, which may unintentionally hide a variability issue.


\begin{table}[h!]
\centering
\caption{Comparison of results obtained on our Bovine ECS dataset. For each method, we carried out five evaluations, each time with different training seeds, and we provide the mean and standard deviation for the per-class F1-score, global accuracy, and temporal accuracy. The best scores are highlighted in bold, excluding CLEmbryo\_NTA. CLEmbryo\_NTA corresponds to our method without the temporal averaging process.}
\label{tab:bovine-classif}
\begin{tabular}{cc|cccc|c||c} 
\hline
\multicolumn{2}{l|}{\multirow{2}{*}{\diagbox{Metrics}{Methods}}} & ESOD     & CNNs-CRF & EmbryosFormer & R2D1         & CLEmbryo        & CLEmbryo\_NTA \\ 
\cline{3-8}
\multicolumn{2}{l|}{}                      & VGG-16    & ResNet-50 & ResNet-50   & R(2+1)D-18      & CSN-50         & CSN-50    \\ 
\hline
\multirow{9}{*}{F1-score per class} & 1             & 85.00 ± 3.34 & 89.26 ± 1.38   & 85.09 ± 1.70 & 92.10 ± 1.59     & \textbf{93.67 ± 0.58} & 93.31 ± 0.66       \\
                  & 2             & 70.25 ± 3.46 & 75.92 ± 1.09    & 65.67 ± 1.10 & 83.82 ± 1.99     & \textbf{86.67 ± 0.77} & 86.35 ± 0.65       \\
                  & 3             & 0 ± 0     & 15.68 ± 4.07    & 3.41 ± 4.67  & 17.45 ± 9.91     & \textbf{37.94 ± 5.23} & 36.92 ± 4.88       \\
                  & 4             & 42.89 ± 24.46 & 57.87 ± 1.30     & 58.83 ± 1.22 & 73.90 ± 1.34     & \textbf{76.64 ± 1.77} & 76.12 ± 2.03       \\
                  & 5             & 0 ± 0     & 4.98 ± 4.48     & 0 ± 0     & 7.12 ± 7.08      & \textbf{19.76 ± 2.14}  & 18.90 ± 1.47       \\
                  & 6             & 0 ± 0     & 5.26 ± 1.01     & 0 ± 0     & 13.35 ± 7.32     & \textbf{15.15 ± 7.43} & 12.91 ± 7.49       \\
                  & 7             & 0 ± 0     & 2.87 ± 2.97     & 0 ± 0     & 7.36 ± 4.52      & \textbf{28.38 ± 6.04} & 28.19 ± 5.45       \\
                  & 8             & 71.48 ± 4.04 & 72.43 ± 2.10    & 73.14 ± 2.07 & \textbf{85.97 ± 1.45} & 85.46 ± 1.43      & 85.38 ± 1.20       \\
                  & 9+             & 0 ± 0     & 31.90 ± 4.46    & 45.68 ± 2.62 & 70.80 ± 6.01 & \textbf{71.96 ± 2.42}      & 71.66 ± 1.67       \\ 
\hline
\multicolumn{2}{c|}{Accuracy ($\vartheta$)}           & 65.52 ± 4.43 & 65.30 ± 1.44    & 66.62 ± 1.12 & 79.95 ± 0.62     & \textbf{81.25 ± 0.89} & 80.88 ± 0.73       \\ 
\hline
\multicolumn{2}{c|}{Temporal accuracy ($\zeta$)}         & 27.14 ± 3.94 & 26.89 ± 2.52    & 18.95 ± 1.83 & 38.82 ± 3.07     & \textbf{61.35 ± 3.43} & 59.06 ± 2.90       \\
\hline
\end{tabular}
\label{tab:bovine-classif}
\end{table}

\begin{table}[h!]
\centering
\caption{Comparison of results obtained on the Mouse Embryos dataset\cite{mouse-dataset}. For each method, we carried out five evaluations, each time with different training seeds, and we provide the mean and standard deviation for the per-class F1-score, global accuracy, and temporal accuracy. The best scores are highlighted in bold, excluding CLEmbryo\_NTA. CLEmbryo\_NTA corresponds to our method without the temporal averaging process.}
\begin{tabular}{cc|cccc|c||c} 
\hline
\multicolumn{2}{l|}{\multirow{2}{*}{\diagbox{Metrics}{Method}}} & ESOD     & CNNs-CRF    & EmbryosFormer & R2D1     & CLEmbryo        & CLEmbryo\_NTA   \\ 
\cline{3-8}
\multicolumn{2}{l|}{}                      & VGG-16    & ResNet-50   & ResNet-50   & R(2+1)D-18  & CSN-50         & CSN-50     \\ 
\hline
\multirow{7}{*}{F1-score per class} & 1             & 98.26 ± 1.08 & 98.66 ± 2.08  & 98.43 ± 0.59 & 95.65 ± 1.34 & \textbf{99.61 ± 0.04} &  99.48 ± 0.04 \\
                  & 2             & 95.52 ±2.66 & 98.33± 0.72  & 98.16 ± 0.55 & 97.43 ± 0.97 & \textbf{99.58 ± 0.10} & 99.62 ± 0.08 \\
                  & 3             & 0 ± 0    & 24.05 ± 37.38 & 70.68 ± 6.99 & 34.19 ± 23.97 & \textbf{90.20 ± 3.40} &  91.29 ± 3.38 \\
                  & 4             & 86.18 ± 4.33 & 90.44 ± 2.50  & 94.58 ± 1.31 & 94.14 ± 1.04 & \textbf{98.97 ± 0.87} &  99.40 ± 0.28 \\
                  & 5             & 0 ± 0    & 6.26 ± 10.86  & 28.62 ± 5.64 & 14.37 ± 11.17 & \textbf{70.36 ± 6.16} &  70.32 ± 4.97 \\
                  & 6             & 8.71 ± 19.48 & 6.19 ± 11.54  & 45.56 ± 8.85 & 19.56 ± 3.76 & \textbf{69.32 ± 10.54} &  57.79 ± 15.84 \\
                  & 7             & 0 ± 0    & 0.51 ± 1.02  & 61.48 ± 0.51 & 40.52 ± 20.76 & \textbf{72.71 ± 5.51} & 52.84 ± 7.61 \\ 
\hline
\multicolumn{2}{c|}{Accuracy ($\vartheta$)}                  & 89.57 ± 2.67 & 92.24 ± 1.51  & 94.54 ± 0.71 & 91.98 ± 0.88 & \textbf{97.73 ± 0.61} & 98.39 ± 0.52 \\ 
\hline
\multicolumn{2}{c|}{Temporal accuracy ($\zeta$)}             & 2.91 ± 3.61 & 38.07 ± 17.43 & 69.69 ± 1.69 & 43.30 ± 9.34 & \textbf{85.14 ± 3.00} & 84.00 ± 3.99 \\
\hline
\end{tabular}
\label{tab:mouse-classif}
\end{table}

\subsection*{Pretraining}

\begin{table}[h!]
\centering
\caption{Results on our Bovine ECS (respectively Mouse Embryos) dataset, in terms of per-class F1-score, global accuracy, and temporal accuracy, obtained by our CLEmbryo method pre-trained on the Mouse Embryos (respectively Bovine Embryos) dataset. we carried out five evaluations each time with different training seeds, and we provide the mean and standard deviation. The best scores are highlighted in bold.}
\begin{tabular}{cc|cc|cc} 
\hline
\multicolumn{2}{l|}{\multirow{2}{*}{\diagbox{Metrics}{Pretraining}}} & \multicolumn{2}{c|}{Bovine}          & \multicolumn{2}{c}{Mouse}            \\ 
\cline{3-6}
\multicolumn{2}{l|}{}                        & No pretraining     & Pretraining      & No pretraining     & Pretraining      \\ 
\hline
\multirow{9}{*}{F1-score per class} & 1               & 93.67 ± 0.58      & \textbf{94.07 ± 0.12} & 99.61 ± 0.04      & \textbf{99.77 ± 0.08} \\
                  & 2               & 86.67 ± 0.77 & \textbf{86.80 ± 0.65}     & 99.58 ± 0.10      & \textbf{99.80 ± 0.07} \\
                  & 3               & 37.94 ± 5.23      & \textbf{41.30 ± 6.61} & 90.20 ± 3.40      & \textbf{94.55 ± 2.25} \\
                  & 4               & \textbf{76.64 ± 1.77}      & 76.02 ± 1.12 & 98.97 ± 0.87      & \textbf{99.00 ± 0.93} \\
                  & 5               & \textbf{19.76 ± 2.14}      & 14.58 ± 5.55 & 70.36 ± 6.16 & \textbf{76.34 ± 3.30}      \\
                  & 6               & 15.15 ± 7.43 & \textbf{21.55 ± 3.89}     & \textbf{69.32 ± 10.54} & 62.81 ± 10.15      \\
                  & 7               & 28.38 ± 6.04      & \textbf{28.99 ± 4.09} & \textbf{72.71 ± 5.51}      & 64.20 ± 4.24 \\
                  & 8               & 85.46 ± 1.43      & \textbf{85.67 ± 1.38} &            &            \\
                  & 9+               & 71.96 ± 2.42      & \textbf{72.85 ± 3.39} &            &            \\ 
\hline
\multicolumn{2}{c|}{Accuracy ($\vartheta$)}                    & 81.25 ± 0.89      & \textbf{81.46 ± 0.70} & \textbf{97.73 ± 0.61}      & 97.30 ± 0.15 \\ 
\hline
\multicolumn{2}{c|}{Temporal accuracy ($\zeta$)}                & 61.35 ± 3.43      & \textbf{62.44 ± 1.48} & 85.14 ± 3.00      & \textbf{89.36 ± 1.05} \\
\hline
\end{tabular}
\label{tab:pretraining}
\end{table}

Finally, we tested if pretraining our method on the Mouse Embryos (respectively Bovine ECS) dataset could be beneficial on the Bovine ECS (respectively Mouse Embryos) dataset. Although the Bovine Embryos dataset is more difficult to process than its mouse counterpart, both datasets show similar events and patterns. It is now well-known that pretraining a deep learning model on another dataset is usually beneficial \cite{pretraining}. Results are reported in Table \ref{tab:pretraining}. Pretraining improves overall temporal accuracy. On the bovine dataset, gains are modest yet consistent across almost all evaluation metrics. On the mouse dataset, it enhances early-stage performance but degrades predictions on late stages. This may be due to species-specific features and transfer biases, possibly amplified by the higher complexity of the bovine data used for pretraining.

\section*{Discussion}
\hypertarget{sec:discussion}{}
We developed a novel method for predicting the cell stages of animal embryos and assessed its performance 
on embryo video datasets for two different species: bovine and mouse\cite{mouse-dataset}. We also elaborated a video dataset of annotated bovine embryos. 
This bovine biological material is particularly challenging because of the darkness of the cells due to the presence of high quantities of intracellular lipids\cite{Hirotada2001},\cite{Abe2002},\cite{GENICOT20051181}. This results in a lack of detail in the bovine embryo cells compared to mouse and human embryos that present a higher degree of transparency.

In summary, we formulated the problem as a supervised multi-stage classification while overcoming imbalanced data distribution.
Our experiments demonstrated the interest of introducing, for this classification task, supervised contrastive learning, focal loss training, and the lightweight 3D convolutional network CSN-50 as subsequence encoder. 
Our CLEmbryo model generalizes well, after retraining on the new dataset. Indeed, CLEmbryo exhibits high performance in two mammalian datasets. CLEmbryo was able to produce convincing results on embryos from different animal species and on videos acquired with different setups.

We favorably compared our CLEmbryo method with other state-of-the-art methods on both the Bovine ECS dataset and the NYU Mouse Embryos dataset. CLEmbryo provides the best scores for almost all of the evaluation metrics, while remaining lightweight. This allowed us to efficiently and accurately predict the cell stages of animal embryos.
The improved performance of CLEmbryo is first due to the supervised contrastive learning framework. Adding this auxiliary head to the network to bring the embedding vectors of the same class of images closer together, results in a more robust and accurate model. The focal loss allowed us to improve the results on the intermediate classes, thereby increasing the temporal accuracy of our method. 
The choice of the network architecture is also impactful.
Unlike LSTM used in\cite{stage-detection-synergic}, a 3D-CNN, as CSN-50, focuses on local temporal view of the input sequence\cite{lstm-3dcnn}, which allows it to better capture short intervals of intermediate cell stages. The LSTM architecture is more sensitive to this class imbalance, since it leads to a more global temporal view of the input, and therefore, is unable to focus on the short appearance of intermediate classes.

As far as the temporal context is concerned, we have the following observations. EmbryosFormer takes advantage of the full temporal context. Although effective on the simpler mouse dataset, its performance drops on the more complex bovine one, becoming comparable to ESOD or CNNs-CRF. This may be due to its reliance on global temporal patterns, which can be disrupted by missing stages or atypical progressions. In contrast, models using shorter sequences, such as CLEmbryo and R2D1, perform better, likely benefiting from more stable local dynamics. CNNs-CFR, which takes one frame in the main path and two consecutive frames in the temporal path performs less well, possibly due to the temporal context being too short. ESOD uses 32-frame sequences with an LSTM to model temporal patterns. As explained previously, LSTM works correctly for predominant developmental stages, but almost completely fails for intermediate stages.
On our side, CLEmbryo does not require the integration of the monotonic growth constraint.
to get better results than the other methods. This makes our method more flexible. However, as demonstrated in Table \ref{tab:seq-length} of the ablation study, having a sufficiently long temporal context is also a key to achieving good performance. In addition, our method provides results with low variability. This property of stability is of fundamental importance to biologists.


Although CLEmbryo performs very well on the Bovine ECS dataset, there is still room for extension. 
It is now known that IVP embryos can present variable cleavage durations and cleavage errors such as direct cleavage, i.e., division of a cell directly into three or four daughter cells, or reverse cleavage, i.e., cleavage followed by fusion of the daughter cells. These variations have a direct influence on embryo viability. Consequently, it is key to determine the number of cells in an embryo at any given time.
The CLEmbryo correctly manages normal and direct cleavage but has not yet been tested on reverse cleavage. It should be able to handle it since it does not involve the monotonic growth constraint. The frequency of reverse cleavage is not widely documented but may reach up to 25\% of transferable embryos in some studies and may reduce the quality of these embryos\cite{Jin2022},\cite{Sugimura2025}. Therefore, it deserves to be addressed in future work. This is challenging because reverse cleavage embryos have unstable developmental dynamics; after cell division, the number of cells can increase and then decrease to the initial number due to the fusion of two daughter cells.
Consequently, our future work will also involve the extension of our dataset of bovine embryo videos to include examples of such cleavage behavior.
More research is also needed to determine the exact timing of cell divisions. Cell division may last several minutes and be documented through successive frames.
The transition from one cell stage label to the next between two frames is not sufficient to infer the precise timing of the start and end of cell division. This would require going beyond the cell stage classification process.


\section*{Data availability}
The Mouse Embryos dataset was introduced in\cite{mouse-dataset} and is publicly available. Our Bovine ECS dataset is available for research
purpose only, upon request to INRAE by email to Alline de Paula Reis at 
\href{mailto:alline.reis@vet-alfort.fr}{\tt alline.reis@vet-alfort.fr}.

\bibliography{sample}

\section*{Acknowledgements}
The authors would like to acknowledge the collaboration of Dr. Véronique Duranthon and Brigitte Marquant-LeGuienne for the experimental protocol design for the embryo production.
The production of the original embryo data was funded by CRB-Anim.
Yasmine Hachani's doctoral fellowship is funded by Inria-INRAE. Operation of the research project is also partly funded by the DIGIT-BIO program of INRAE.

\section*{Author contributions statement}
Y.H., P.B., E.F., and A.D.P.R. conceived this research work and the experiments. Y.H. implemented the software, ran the experiments. A.D.P.R., S.R and L.L produced the bovine embryos and acquired the videos with the Primovision system. A.D.P.R. annotated the videos. Y.H., P.B., E.F., and A.D.P.R wrote the paper. Y.H. drew the figures.

\section*{Additional Information}

\section*{Competing interests}
The authors declare no competing interests.

\section*{Approval for animal experiments}
This research study was conducted using data available in our laboratory. No live animals or euthanised animals were used to create the original data. The semen was acquired from a commercial company and the cumulus oocyte complexes were harvested from ovaries recovered \textit{post-mortem} in a commercial slaughterhouse. Both these companies and our laboratory are based in France and state-approved. The necessary authorisations for the use of \textit{post-mortem} biological material have been obtained from the responsible Ministry.

\end{document}